# First principles modeling of defects in the $Al_2O_3$/$In_{0.53}Ga_{0.47}As$ system

Gabriel Greene-Diniz[1], Kelin J. Kuhn[2], Paul K. Hurley[1], and James C. Greer[1]

[1]*Tyndall National Institute, Lee Maltings, Prospect Row, Cork, Ireland*

[2] *Department of Materials Science and Engineering, Cornell University, Ithaca, NY, USA 14853-2201*

Density functional theory paired with a first order many-body perturbation theory correction is applied to determine formation energies and charge transition energies for point defects in bulk $In_{0.53}Ga_{0.47}As$ and for models of the $In_{0.53}Ga_{0.47}As$/$Al_2O_3$ interface. The results are consistent with previous computational studies that $As_{Ga}$ antisites are candidates for defects observed in capacitance voltage measurements on metal-oxide-semiconductor capacitors, as the $As_{Ga}$ antisite introduces energy states near the valence band maximum and near the middle of the energy band gap. However, substantial broadening in the distribution of the $Ga_{As}$ charge transition levels due to the variation in the local chemical environment resulting from alloying on the cation (In/Ga) sublattice is found, whereas this effect is absent for $As_{Ga}$ antisites. Also, charge transition energy levels are found to vary based on proximity to the semiconductor/oxide interface. The combined effects of alloy- and proximity-shift on the $Ga_{As}$ antisite charge transition energies are consistent with the distribution of interface defect levels between the valence band edge and midgap as extracted from electrical characterization data. Hence, kinetic growth conditions leading to a high density of either $Ga_{As}$ or $As_{Ga}$ antisites near the $In_{0.53}Ga_{0.47}As$/$Al_2O_3$ interface are both consistent with defect energy levels at or below midgap.

## I. INTRODUCTION

As the number of transistors on a single integrated circuit has increased to the point of exceeding tens of billions for logic circuits and hundreds of billions for random access memories, critical dimensions of 10 nm or below are required for production technologies. The negative impact that electrically active defects have on device performance for larger device scales is exacerbated for nanoscale transistors; conventional problems such as gate screening due to interface and oxide charges occur to an even higher degree, and larger device to device variations are introduced that must be understood and eliminated, or at least controlled [1,2,3]. Nanoelectronics design finds itself in a regime of power-constrained scaling in which power density cannot be significantly increased in silicon technologies without overcoming problems related to further device scaling [4,5]. Due to the larger electron mobility compared to silicon, III-V semiconductors such as $In_{0.53}Ga_{0.47}As$ remain candidates for overcoming barriers to scaling into the sub 10 nm device technology nodes [5] and for high performance applications. It should be noted that the bulk mobility for devices with channel lengths less than the electron mean free path display ballistic transport and the bulk mobility becomes less of a criterion. For ballistic transport, the current drive is determined by the source density of states and the electron injection velocity. Due to a low electronic density of states (DoS) at the conduction band edge for typical III-V materials considered for electronics, the advantage of a 'high mobility' material is no longer a decisive factor for scaled devices. The low density of states limiting current drive in ultra-scaled devices is referred to as the DoS bottleneck.

A large defect concentration at III-V/oxide interfaces has traditionally been another significant obstacle to the integration of III-V materials into mainstream complimentary metal-oxide-semiconductor (CMOS) technologies as the electrostatic control of the semiconducting channel's charge density at the semiconductor surface by the gate electrode becomes screened by charged defect states. Determination of the atomic structure of electrically active defects aids developing means to either eliminate the formation of the defects, or to devise schemes to passivate the defects subsequent to their formation. Recent electrical studies on metal-oxide-semiconductor (MOS) capacitors consisting of $In_{0.53}Ga_{0.47}As$/$Al_2O_3$ and $In_{0.53}Ga_{0.47}As$/$HfO_2$ allow for interface density of states ($D_{it}$) in the energy gap of $In_{0.53}Ga_{0.47}As$ near the semiconductor/oxide interface to be extracted [6-9]. Key findings include the following. The dominant interface defects are electrically active for the range of gate voltages typical for device for operation; the defects are believed to be associated with the semiconductor and largely independent of the specific gate oxide material; the highest defect density within the band gap occurs between the VBM and midgap [6]. The electrical techniques applied to extract the defect states in the band gap cannot directly identify the atomic structure of the defects, nor provide insight into avoiding formation of the defects, nor guide strategies for defect passivation.

Hence the purpose of the calculations presented in this study is to narrow the possible set of atomistic configurations giving rise to defects levels in the band gap, and thereby motivate the development of growth conditions and processing steps that either passivate the defects or avoid their formation.

Recent density functional theory (DFT) investigations of defects in related III-V (GaAs, InAs, and InGaAs alloys) materials have been reported. These studies provide predictions for the stability and amphoteric nature of native defects in bulk and oxide-terminated surface models using hybrid exchange-correlation (XC) functionals [10-14], with the hybrid functionals chosen to overcome the DFT band gap problem. In such approaches a fraction $a$ of Hartree-Fock exchange is mixed into the XC functional (giving rise to a hybrid functional), and the value of $a$ is usually adjusted to reproduce the experimental band gap. These authors [10-14] conclude that the position of the midgap charge transition levels (CTLs) of the $As_{Ga}$ antisite, combined with a predicted lower formation energy relative to other commonly studied point defects, , suggest that this antisite is responsible for the midgap $D_{it}$ states observed in the capacitance-voltage (CV) response of $In_{0.53}Ga_{0.47}As$/high-$k$ oxide MOS capacitors. In other works, studies of bonding mechanisms at the III-V/oxide interfaces $GaAs/Al_2O_3$ and $GaAs/HfO_2$ have been presented [15], and predictions of charge transition levels (CTLs) of As and P vacancies at (110) oriented GaAs and InP surfaces are reported in ref. [16]. The latter utilizes many body perturbation theory (MBPT) to avoid the need to empirically parameterize the XC functional. The approach to determining CTLs presented in ref. [16] is applied in the present study.

Comparison of the formation energetics and CTLs of defects yielding midgap $D_{it}$ states occurring in a bulk-like environment with defects occurring in the vicinity of the semiconductor/oxide interface has not been performed to date. To explore the influence of the position of the defect with respect to the semiconductor/oxide interface, a comparison is performed using a 64 atom simulation cell with periodic boundary conditions to model bulk $In_{0.53}Ga_{0.47}As$ with 17 In and 15 Ga atoms randomly distributed on the cation sublattice; further details are given in section II.B. This structure is chosen as a reference configuration for defects forming in the "bulk". Three point defects are studied with the bulk simulation cell: the antisites $Ga_{As}$ (Ga on an As site) and $As_{Ga}$ (As on a Ga site), and a Ga vacancy, the latter denoted as $V_{Ga}$. Only point defects in the semiconducting region are chosen due to the experimental indication that the measured density of interface defects is largely independent of the oxide [8,17,18]. This new set of calculations allows for a comparison between the use of hybrid functionals and the DFT+$GW$ approach described in section II.A. The DFT+$GW$ approach allows for calculations free of the empirical parameterization introduced to calibrate the theoretical band gap to the experimental value as required with hybrid functional approaches. Additionally, the explicit influence of alloying on the cation sublattice on the $Ga_{As}$ CTL is investigated. To explore the effects of the chemical environment on the antisites $Ga_{As}$ and $As_{Ga}$ that form within the vicinity of a semiconductor/oxide interface, an interface model is introduced with $Al_2O_3$ passivation of a (100) $In_{0.53}Ga_{0.47}As$ surface. The CTLs of these point defects are re-examined for varying positions in the semiconducting region relative to the interface. The use of the interface model also permits the As dimer ($As_2$) to be studied thereby providing another reference point for the calculations, as this surface defect has been studied in a similar $InGaAs/Al_2O_3$ interface model using hybrid DFT [14]. A surface cleaved at an arsenic layer of atoms is used in building the interface model to reflect that for growth of indium gallium arsenide by metal organic chemical vapor deposition (MOCVD) that the samples are typically cooled in the presence of an $AsH_3$ flux.

Defect formation energies have been studied as a function of chemical potential to assess the stability of defects in anion-rich, cation-rich, or varying stoichiometric conditions [13,19,20] for defects in a bulk-like environment. CTLs calculated by evaluating changes to formation energies have been studied for bulk defects [13] and for the arsenic dimer ($As_2$) which can only be constructed from an explicit interface model [14]. Key results from these calculations are summarized in fig. 1. This work focuses on the evaluation of CTLs and extends previous studies by considering defects in a bulk-like region and in regions only a few atomic distances from an explicit semiconductor/oxide interface, as well as including the effects of alloying on the cation sublattice. DFT+$GW$ is employed to aid in reducing errors when determining electron affinities and ionization potentials required to add or remove charges to the defect sites. In our calculations the following broadening mechanisms are taken into account: the broadening of energy levels due to proximity to the interface, the local environment due to alloying, and the effects of thermal broadening on the distribution of defect states. These considerations allow for an analysis that can be directly compared to experimental capacitance-voltage measurements. This will be shown to lead to conclusions consistent with previous studies but also points to additional defects that can give rise to experimentally determined interface defect states.

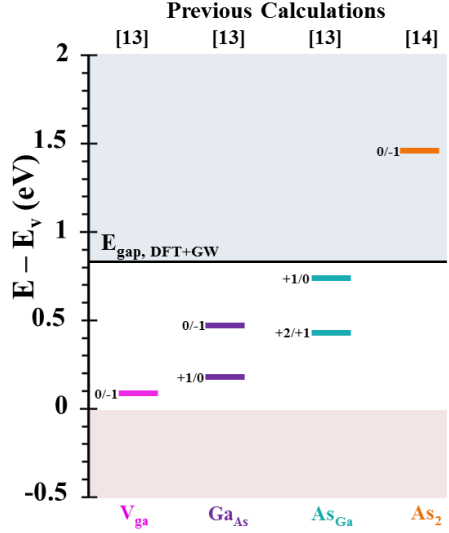

Figure 1 - A summary of previous calculations of charge transition levels employing hybrid density functional theory for bulk defects [13] and the interface defect $As_2$ [14].

## II. METHODS

This section describes the methods for calculation of the formation energies and CTLs of defects in the bulk and at interfaces using models developed for $In_{0.53}Ga_{0.47}As$ bulk and the $In_{0.53}Ga_{0.47}As/Al_2O_3$ interface. Formation energies are given as the energy required to form a defect relative to the energy of a defect-free supercell achieved by adding and/or removing atoms to and/or from an ideal atom reservoir. All energies are for configurations relaxed to minimize the total cell energy. CTLs occur at the value of the Fermi level for which the charge state of a defect changes formally by one electronic charge. Charged state formation energies are decomposed into two contributions following the approach described in ref. [16]: a structural relaxation energy term $\Delta$, and an electron addition (removal) term $A$ ($I$), given by the vertical electron affinities and ionization potentials, respectively. A DFT approximation is retained for the relaxation term $\Delta$ which is found as the energy difference between two supercells with the same number of electrons thereby avoiding the difficulties commonly associated with approximate XC functionals when treating systems with varying numbers of electrons [21,22]. The $GW$ approximation as a quasiparticle theory is well equipped to treat addition or removal of charges into a system. The more computationally demanding $GW$ approximation is reserved for the calculation of electron affinities $A$ and ionization potentials $I$.

The DFT calculations are performed using norm-conserving pseudo-potentials and the Perdew-Zunger form of the local density approximation (LDA) to the XC functional [23] to determine total energies for the relaxed supercell configurations. The relaxations of the atomic positions within a supercell and atomic reference energies are calculated with Quantum Espresso [24]. Relaxation of the lattice vectors for the defect free cell circumvents any possible errors due to non-equilibrium cell volumes or lack of convergence with respect to energies as discussed in ref [19]. Following the structural relaxation of the pristine cell, point defects are inserted and the atomic positions within the supercell are allowed to relax to a new minimum energy configuration. This procedure is also repeated for each charge state of a defect with a compensating uniform background charge to prevent divergences of the Coulomb interactions between periodic image charges. Regarding the numerical details of the DFT calculations and subsequent $GW$ corrections, the following parameters were applied to pristine and defect simulation cells: a 60 Rydberg kinetic energy cutoff, 2x2x2 (2x2x1) $k$-point meshes for geometry optimization of the bulk (surface) defects, and 1772 unoccupied bands along with $\sqrt{N} = 5041$ where N is the number of elements in the dielectric matrix is used to calculate the self-energy required for the $GW$ correction. This value of N was chosen based on the convergence of the band gap of the pristine cell. A bulk band gap of 0.84 eV for the pristine cell is obtained from a $GW$ calculation, in good agreement with the measured low temperature band gap of 0.82 eV [25].

For the case of bulk defects, a 64 atom supercell with 17 In and 15 Ga atoms randomly distributed on the cation sublattice which hosts a representative distribution for arsenic sites with a varying number of In/Ga nearest neighbors is selected; the distribution of nearest neighbors about the arsenic sites is shown in fig. 2. The slightly asymmetric distribution about $N_{As-In}$ = 2 is consistent with the stoichiometry of the cation sublattice in $In_{0.53}Ga_{0.47}As$. Defects are introduced into these cells, and formation energies and structural relaxation energies $\Delta$ are calculated. The *A* addition and removal *I* energies are calculated from the *GW* approximation using the relaxed geometries obtained from DFT.

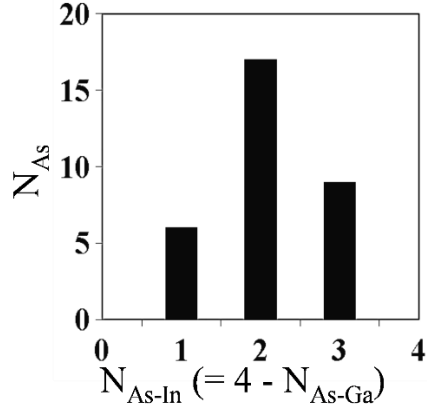

Figure 2 - Histogram associated with the number of cations bonding directly to each As atom in the 64 atom bulk cell. $N_{As}$ is the number of As atoms with $N_{As-In}$ bonds to indium atoms.

For the surface model, passivation $Al_2O_3$ is chosen to compare to recent experimental data [8]. The building of interfacial models to represent oxides on III-V surfaces requires more advanced considerations to achieve passivation relative to $Si/SiO_x$ interfaces [26]. The interface model proposed by Robertson and Lin for $Al_2O_3$ passivation [15] is used; this model satisfies electron-counting rules and results in a stable, intrinsic (band gap is free of defect states and the Fermi level lies at midgap) semiconducting region. The semiconducting region bonding directly to the oxide consists of an As layer. The bottom of the slab is an As-terminated surface passivated by pseudo-hydrogens with charges $q = 0.75$. For the structural relaxation of the point defects in $Al_2O_3$ passivated (100)-surface $In_{0.53}Ga_{0.47}As$ models, 304 atoms (208 semiconductor atoms, 40 oxide atoms, 24 H and 32 pseudo-H) supercells are used resulting in the periodic images for the defects being separated by approximately 16 Å (the supercell size is doubled in the *x* and *y* directions compared to that shown in fig. 3). Defects are introduced into the model and formation energies calculated. Periodic images of the slabs in the direction normal to the interface are separated by ~18 Å of vacuum. These distances minimize the interaction between periodic images of defects while maintaining a reasonable computational effort. For the distribution of In and Ga atoms in the surface model we investigate the use of special quasi-random structures (SQS) [27,28] to define the cation sublattice. While the SQS described in ref. [27] is constructed to mimic the multi-site correlation functions of a bulk fcc $A_{0.5}B_{0.5}$ alloy, application of the SQS8 configuration to the 304 atom (100)-surface model results in a composition of $In_{0.53}Ga_{0.47}$ for the cation sublattice due to the truncation of the structure in the slab model; the resulting cation configuration is representative of a random $In_xGa_{1-x}$ alloy with a slight excess of In content as reflected in the histogram of As-In/Ga bonds shown in fig. 4. As in the 64-atom bulk case, the small asymmetry about $N_{As-In}$ = 2 is consistent with a small excess of In content on the cation sublattice. Thus, both the surface model and bulk model used in this work adopt cation configurations that are both representative of random alloys and maintain compositions consistent with each other, as well as with recent experimental studies [8].

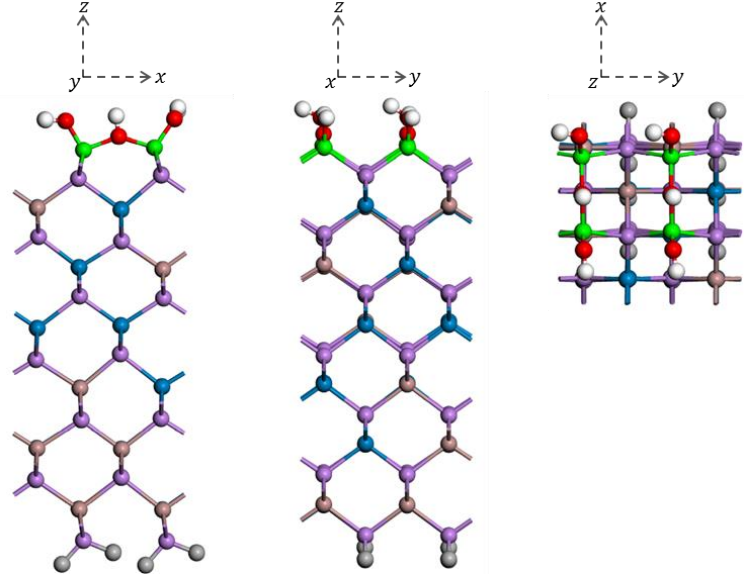

Figure 3 - 1x1 unit cells of (100) InGaAs passivated with $Al_2O_3$. $x$, $y$ and $z$ supercell directions are shown at the top. White, green, and red spheres indicate H, O, and Al. Pink, blue and brown spheres indicate As, In, and Ga. Grey spheres on the bottom are for pseudo-hydrogens with valence = 0.75.

Due to the computational demand, smaller 152 atom supercells were used to compute the *GW* correction; smaller supercells for the electron addition energy term has been used previously when employing the *GW* correction for both neutral and charged cells [16]. The justification for the use of smaller cells hinges on the lack of electrostatic (Hartree) contributions to the self-energy correction $\Sigma$, *i.e.* the latter only involves exchange and correlation terms leading to the supercell dimensions having a less pronounced effect on $\Sigma$.

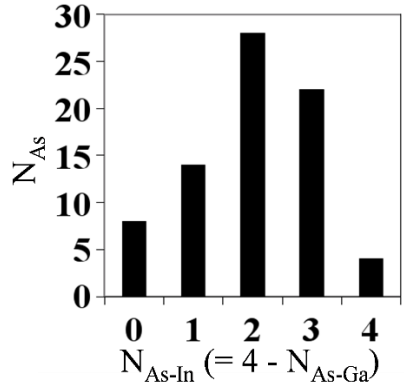

Figure 4 - Histogram associated with the number of cations bonding directly to each As atom which is 4-fold coordinated to the native cation sub-lattice in the 304 atom surface model. $N_{As}$ is the number of As atoms with $N_{As-In}$ bonds to indium atoms.

Defect formation energies are calculated as

$$E_{form}(D) = E(D) - E(ref) + \sum_a n_a E(a), \quad (1)$$

where $E(D)$ is the total energy of a neutral relaxed simulation cell containing a single defect, $E(ref)$ is the energy of a neutral relaxed simulation cell of the pristine host, $E(a)$ is the energy of an isolated atom, and $n_a = +1$ (-1) if atomic species $a$ is removed (added). The latter term corresponds to the chemical potentials of added or removed species, *i.e.* the energy relative to an ideal atomic reservoir. This is analogous to the standard procedure for calculating interfacial energies, in which grand canonical thermodynamics are used to evaluate energies of surfaces and interfaces relative to the dissociated components

[29,30].

The thermodynamic CTL $\varepsilon_{q/q'}$ is defined as the value of the Fermi level where energetically competing charge states have equal charge formation energies. The energy for forming a charged defect state $q$, relative to charge state $q'$ is given by

$$E_{form}(q/q') = E(q, R_q) - E(q', R_{q'}) + (q - q')\Delta\varepsilon_F, \quad (2)$$

where $E(q, R_q)$ is the total energy of a simulation cell containing a single defect in charge state $q$ with the geometry relaxed to the configuration of the $q^{th}$ charge state denoted by $R_q$. The final term $\Delta q \Delta\varepsilon_F$ is the Fermi level with respect to the valence band maximum (VBM) and accounts for the transfer of electrons to and from a charge reservoir. The slope of $E_{form}(q/q')$ is given by the coefficient of the $\Delta\varepsilon_F$ term [31], *i.e.* the final change in the formal charge state of a defect after an addition/removal process as the Fermi level is swept from the VBM to the conduction band minimum (CBM). This dependence of $E_{form}$ on the Fermi level can be used to extract the CTL $\varepsilon_{q/q'}$. The energy $E_{form}$ is plotted as a function of $\Delta\varepsilon_F$ for differing charge states, the Fermi level at which differing $E_{form}$ intersect corresponds to a CTL [19,32].

As the name suggests, charge transition levels involve calculations with different numbers of electrons at or near a defect site. Thus the well-known deficiencies of DFT rooted in the lack of a derivative discontinuity in the XC functional with respect to particle number [21,22] exacerbate the computation of CTLs within a DFT framework. For example, the underestimation of the fundamental gap which arises from the above-mentioned weaknesses can lead to a qualitatively wrong picture in which CTLs resonate with the host bands whereas experiment, and more rigorous theoretical approaches would indicate they lie within a semiconductor's band gap [33]. To overcome the shortcomings of DFT, the formation of charged defects can be decomposed into two contributions: the energetic cost of adding or removing an electron, and the energy change arising from the relaxation of surrounding atoms upon addition of the electron or hole. This is achieved by rewriting eq. 2. Let $E(q, R_{q'})$ be the energy of a simulation cell with charge state $q$ but with atomic positions optimized for charge state $q'$. Adding and subtracting $E(q, R_{q'})$ and grouping all terms appropriately yields

$$E_{form}(q/q') = E(q, R_{q'}) - E(q', R_{q'}) + E(q, R_q) - E(q, R_{q'}) + (q - q')\Delta\varepsilon_F$$
$$= A(q, q', R_{q'}) + \Delta(R_q, R_{q'}, q) + (q - q')\Delta\varepsilon_F. \quad (3)$$

The first term on the right-hand-side $A$ represents the energy to add an electron neglecting the rearrangement of surrounding atoms $E(q, R_{q'}) - E(q', R_{q'})$, known as the vertical electron affinity. The second term $\Delta$ arises from the relaxation of atoms in the presence of the extra charge. The decomposition is such that the charge states relating to the energy terms in $\Delta$ do not differ. As $\Delta$ can be computed without changing the number of electrons, the problem of a discontinuous XC functional with respect to electron number does not apply to the relaxation term [16,33], and approximate DFT can be used. However. by necessity the charge state changes for the calculation of the electron affinity $A$, and the *GW* approximation is a suitable method to determine this term. For electron addition ($q' \rightarrow q' - 1$) the vertical transition corresponds to the electron affinity of the defect; in this case the electron is absorbed from the surrounding reservoir into the defect level [31]. It is also possible to form charge states by removal of electrons. In this case it is assumed that the vertical electron ionization energy ($I$) of $q$ to $q'$ equals the negative of the vertical addition energy ($-A$) of $q'$ to $q$. For electron removal ($q \rightarrow q + 1$), a vertical transition would correspond to the ionization energy of the defect, where the removed electron is transferred *from* the defect level *to* the electron reservoir. The charge state formation energy can thus be re-expressed in terms of the energy to remove an electron plus the energy associated with the subsequent relaxation of atoms due to the hole that is created

$$E_{form}(q'/q) = E(q', R_q) - E(q, R_q) + E(q', R_{q'}) - E(q', R_q) + (q' - q)\Delta\varepsilon_F$$
$$= I(q', q, R_q) + \Delta(R_{q'}, R_q, q') + (q' - q)\Delta\varepsilon_F. \quad (4)$$

The *GW* method as implemented in the YAMBO code [34] is applied to calculate the vertical charge addition and removal energies. In this approach, a first order perturbation correction to the Kohn-Sham eigenvalues $\varepsilon_{KS}$ are obtained to yield quasiparticle levels

$$\varepsilon_{qp} = \varepsilon_{KS} + \langle\varphi_{KS}|(\Sigma - V_{xc})|\varphi_{KS}\rangle \quad (5)$$

where $\varphi_{KS}$ are the Kohn-Sham orbitals assumed to be sufficiently similar to the quasiparticle wavefunctions [35], $\Sigma$ is the electronic self-energy, and $V_{xc}$ is the XC potential (subtracted to avoid double counting between the Kohn-Sham eigenvalue and the self-energy expectation value). The separation of the highest occupied and lowest unoccupied levels at the Γ-point is taken as $A$. In order to calculate this correction, the self-energy is required which in turn requires a convolution of the Green's function $G$ and the screened Coulomb interaction $W$. The former is constructed using wave functions and band energies acquired from a self-consistent DFT calculation [34,35] while the latter depends on the inverse dielectric response function $\varepsilon^{-1}$ that takes into account the local field effects and the dynamics of the screened interaction. In reciprocal space, the relation between the screened Coulomb interaction and $\epsilon^{-1}$ can be written as

$$W_{\mathbf{G},\mathbf{G}'}(\mathbf{q},\omega) = \epsilon_{\mathbf{G},\mathbf{G}'}^{-1}(\mathbf{q},\omega) v(\mathbf{q}+\mathbf{G}') \tag{6}$$

Here, $\mathbf{q}$ is an arbitrary wave vector while $\mathbf{G}$ is a reciprocal lattice vector, and $v(\mathbf{q}+\mathbf{G}')$ is the bare Coulomb interaction. The local fields arise from the off-diagonal $\mathbf{G} \neq \mathbf{G}'$ elements. The connection back to DFT-calculated quantities is made from the relation between the static dielectric function and the non-interacting polarizability $P$ obtained from Kohn-Sham wavefunctions and eigenvalues [35].

The plasmon pole approximation (PPA) [34,36] to describe the frequency dependence of $\epsilon^{-1}$ is made. The PPA assumes the spectral function for the screened interaction to be a single narrow peak in the plasmon energy $E$ ($= \hbar\omega$ where $\omega$ is the plasmon frequency) [36]. If this condition does not hold [37], the dielectric function must be explicitly computed throughout the full frequency axis leading to large increases in computational time and memory requirements. A single peak in the dielectric function is observed for bulk $In_{0.53}Ga_{0.47}As$, as well the surface model exhibits a single peak in the imaginary part of the inverse dielectric response function as shown in fig. 5. This allows for the application of the $GW$ method to supercells containing >150 atoms.

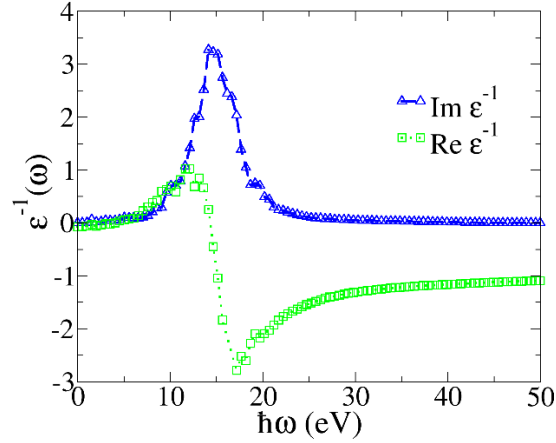

Figure 5 - Real and Imaginary parts at the Γ-point component ($\mathbf{G} = \mathbf{G}' = 0$) of the inverse of the dynamic dielectric matrix, obtained from the $Al_2O_3$ passivated (100) $In_{0.53}Ga_{0.47}As$ surface model, plotted as a function of energy

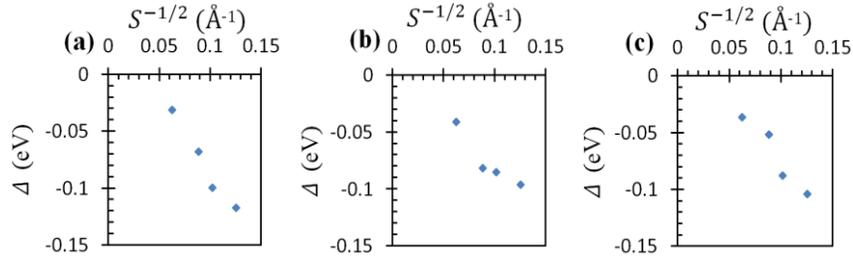

Figure 6 - Energy difference $\Delta = E(q_{-1}, R_{-1}) - E(q_{-1}, R_0)$ for (a), $\Delta = E(q_0, R_0) - E(q_0, R_{-1})$ for (b), and $\Delta = E(q_0, R_0) - E(q_0, R_{+1})$ for (c), all versus $S^{-1/2}$. Variation in $\Delta$ from the smallest to largest supercell is 68 meV for (a), 85 meV for (b), and 55 meV for (c).

Calculations with charged cells have a spurious electrostatic interaction due to the images created by the periodic boundary conditions. For the calculation of the relaxation energies of charged defects, energies are obtained as the difference between energies for simulation cells with the same charge state but in different geometries. Thus errors associated with electrostatic interactions may, to a large degree, systematically cancel using the energy decomposition in eq. 3 [19]. To explore the effect of the error due to interaction between periodic image charges, the $Ga_{As}$ defect in a $q = -1$ charge state is taken as a test case. Fig. 6 shows the relaxation energies $\Delta$ versus the reciprocal of the square root of the surface area $1/\sqrt{S}$ where $S = a \times b$ where $a$ and $b$ are the cell parameters parallel to the plane of the surface in each supercell. Four distinct (100)-surface orientation $In_{0.53}Ga_{0.47}As/Al_2O_3$ models are constructed consisting of 72, 114, 152, and 304 atoms. Figs. 6(a), 6(b), and 6(c) show the variation in the total energy differences with increasing cell size for relaxations from the neutral to negatively charged defect geometry $E(q_{-1}, R_{-1}) - E(q_{-1}, R_0)$ in (a), negative to neutral defect geometry $E(q_0, R_0) - E(q_0, R_{-1})$ in (b), and positive to neutral defect geometry $E(q_0, R_0) - E(q_0, R_{+1})$ in (c). Comparison between the three cases reveals a similar variation of the relaxation energy as a function of surface area for all three cells. Comparison of the charged defect relaxation (a) to the change in the relaxation energy in the neutral cells reveals that the elastic defect-defect interactions dominate the relaxation energies and that the error due to the Coulomb interactions between the charged defect images are substantially less. These results are consistent with the typical estimates for a computational error of ~100 meV in defect formation energies from DFT [14,19,32,33], and comparable to similar errors quoted from experimental studies [7,38] of defect levels. Therefore, as in previous works, the electrostatic correction term in the relaxation energies is omitted due to the favorable cancellation that occurs when taking differences to obtain the formation energies [19]. The resulting estimate of the error in the formation energies due to the finite cell sizes used in these calculations are estimated to be of the order of tens of meV.

## III. RESULTS

The structures for the gallium vacancy $V_{Ga}$, the gallium antisite $Ga_{As}$, and the arsenic $As_{Ga}$ in bulk-like environments following relaxation of the atomic positions and cell parameters in the simulation cells are considered. Similar results for $Ga_{As}$ and $As_{Ga}$ antisites are then presented in section III.B for the surface interface models in the proximity of the $In_{0.53}Ga_{0.47}As/Al_2O_3$ surface, and as well for the arsenic dimer $As_2$ and $Ga_{As}$ when formed directly at the semiconductor/oxide interface are presented.

### A. Bulk defects

#### *1. Structural properties*

Displacement of the atoms neighboring neutral defects are measured relative to a pristine (defect-free) cell, while displacements of the neighboring atoms for the charged defects are measured relative to the relaxed neutral defects. Focusing on the Ga vacancy ($V_{Ga}$), removal of a Ga atom results in the four surrounding As atoms relaxing inward towards the vacancy site on average by 0.5 Å. The tendency of the surrounding anions forming weak bonds across a cation vacancy has been noted in previous computational studies on III-Vs materials [39]. The structural effect of charging the vacancy is slight compared to the structural changes associated with formation of the vacancy, leading to small inward displacements on the order of ~0.01 Å for the four neighboring anions when an electron is added. Removing an electron from the neutral state and allowing the defect site to relax leads to the a positively charged vacancy with in a similar small degree of relative change in the defect configuration but with the four neighboring As atoms becoming displaced outwards away from the vacancy site.

The geometry of the $Ga_{As}$ antisite is studied for the configuration where the antisite bonds with two Ga and two In atoms. Relative to the unrelaxed substitutional site, the bonds between the neutral defect and the surrounding Ga atoms shorten by 0.08 Å, whereas the bonds between the defect and the nearest-neighbor In atoms elongate by 0.04 Å. Thus, the $Ga_{As}$ remains in a four-fold coordinated arrangement but moves slightly away from the tetrahedral symmetry of the non-defective As site. Charging the $Ga_{As}$ site to a positive state pushes the defect center toward the two nearest neighbor Ga atoms by 0.04 Å compared to the position of the neutral defect. Charging this antisite to a negative charge state results in a slight movement of the defect center by 0.03 Å compared to the neutral position but in the opposing direction to the positive state relaxation, and toward the two In nearest neighbors.

For the neutral $As_{Ga}$ antisite, the surrounding bonds exhibit larger changes relative to the bond length changes upon formation of $Ga_{As}$: an outward relaxation of 0.16 Å relative to the bonds of As bonding to Ga. Charging to a positive state causes the As-$As_{Ga}$ bonds to contract by 0.05 Å relative to the neutral case, with tetrahedral symmetry preserved and the defect center remains in the same position relative the relaxed neutral charge state. This displacement is larger for a doubly charged state of the $As_{Ga}$ antisite; an average of 0.06 Å contraction of the anion antisite bond lengths relative to the +1 charge state is found.

For all defects studied, the bond lengths located along atomic planes which bisect the spacing between periodic images of the defect center undergo displacements of less than 0.01 Å on average, a measure relative to the corresponding pristine cell that indicates the defect-defect interactions are limited for the selected cell size. This indicates that spurious elastic effects due to periodic boundary conditions are reduced to acceptable levels in our bulk model and is consistent with the finding of Van de Walle and Neugebauer that 64-atom cells are sufficient to estimate the relaxation energy associated with a point defect in bulk semiconductor calculations [19]. Furthermore, as will be seen, the relatively small relaxations of the neighboring atoms upon changing charge state relative to the relaxed neutral defect geometry suggest that the relaxation energies will tend to be small compared to the vertical charge addition and removal energies.

*2. CTLs for bulk defects*

Bulk CTLs for As and Ga antisites ($As_{Ga}$, $Ga_{As}$) and the As and Ga vacancies ($V_{As}$, $V_{Ga}$) calculated with hybrid DFT have been reported [13]. For the $V_{Ga}$, a neutral to negative transition denoted as $\varepsilon_{0/-1}$, is found to occur at an energy 0.08 eV above the VBM. A $\varepsilon_{+1/0}$ transition for the $As_{Ga}$ antisite occurs at 0.74 eV above the VBM, while the $\varepsilon_{+2/+1}$ transition for the $As_{Ga}$ defect lies close to midgap at 0.42 eV above the VBM. The latter is found to be 50 meV lower than the $\varepsilon_{0/-1}$ transition for $Ga_{As}$, while the $\varepsilon_{+1/0}$ $Ga_{As}$ transition lies slightly above the $\varepsilon_{0/-1}$ $V_{Ga}$ transition found from hybrid DFT calculations. In the work reported in ref. [13], a study of the defect formation energies as a function of growth conditions approximated by varying the chemical potentials of the added/removed species is also performed. Taking this study together with the proximity of the $As_{Ga}$ transition to the midgap $D_{it}$ feature measured at InGaAs/oxide interfaces by electrical spectroscopy [6,7,8], those authors assign the $As_{Ga}$ antisite with the transition level $\varepsilon_{+2/+1}$ as a strong candidate for the defect responsible for experimentally extracted midgap defect densities [13]. The $V_{As}$ vacancy is also considered in ref. [13]; it is shown that this defect exhibits a positive to neutral CTL very close to the CBM, and no midgap CTLs. This finding is coupled with the fact of a high cost of formation of $V_{As}$ for substrates grown under As-rich conditions [8] to rule out $V_{As}$ as a candidate for midgap $D_{it}$ [13]. In addition, under (approximated) Ga-rich growth conditions the $Ga_{As}$ antisite is found to be more stable than $V_{As}$ [13], which also suggests the relatively small contribution of $V_{As}$ to midgap $D_{it}$, even for growth conditions corresponding to low As concentration. It should also be noted that $V_{As}$ is situated on an anion site and hence can be subjected to alloying effects due to changes in the cation sublattice. The influence of such effects on the energy of $V_{As}$ has already been studied in detail [40]. Murphy *et al* found a large variation (~500 meV) of the $V_{As}$ formation energy as a function of variations in the alloy. However, the authors of ref. [40] did not consider anion-situated antisites in their work, nor did they report calculations of CTLs as a function of the local alloy. We also investigate the role of local alloying on an anion situated defect, however due to the aforementioned findings regarding the midgap $D_{it}$ candidacy of $V_{As}$, combined with a lack of studies of the effect of alloying on CTLs of antisites, we exclude $V_{As}$ from this work, and instead focus on the effects of changes to the local cation alloy on the $Ga_{As}$ antisite (presented at the end of this section).

The energetics for $V_{Ga}$, $As_{Ga}$ and $Ga_{As}$ defects determined within the DFT+$GW$ approach are computed and compared to the hybrid DFT results. The transition to a neutral $V_{Ga}$ defect is obtained by adding an electron to the relaxed geometry of the positively charged defect and subsequently relaxing the geometry of the simulation cell of the neutral defect. The energy difference between the neutral cell in its relaxed geometry and the neutral cell with the atomic positions fixed to those of the +1 charge state results in a relaxation energy of $\Delta$ = -0.01 eV. The DFT+$GW$ approach yields a vertical electron affinity of 0.07eV which results in a charge state formation energy of $E_{form}(q_0) = 0.06$eV for the neutral $V_{Ga}$ defect. The -1 charge state is formed by adding an electron to the neutral defect. The DFT-calculated relaxation energy plus the $GW$ correction to the vertical electron affinity yields $E_{form}(q_{-1})$ of 0.21 eV for the $V_{Ga}$ defect. As anticipated from the analysis of the charged defect geometries, both charge state relaxation energies for the vacancy are relatively low. The charge formation energies vary as the Fermi level $\varepsilon_F$ is varied and a CTL occurs as the energies cross and a new charge state becomes more stable

$$\varepsilon_{q/q'} = \frac{E_{form}(q') - E_{form}(q)}{q - q'}, \qquad (7)$$

for $V_{Ga}$ $q = 0$ and $q' = -1$, the CTL $\varepsilon_{0/-1}$ occurs at 0.15 eV above the VBM.

For the +1 charge state of the Ga$_{As}$ antisite in the relaxed configuration, an electron affinity of 0.11 eV is obtained for the DFT+$GW$ approximation. Together with the structural relaxation contribution, this yields a formation energy of 0.06 eV for the neutral charge state of Ga$_{As}$. Forming the negative and positive charge states by adding and removing an electron to and from the neutral state, the CTLs $\varepsilon_{+1/0}$ and $\varepsilon_{0/-1}$ are found to lie at 0.19 eV and 0.31 eV above the VBM, respectively.

For the As$_{Ga}$ antisite, the +1 and +2 states are formed by successive removal of electrons starting from $q = 0$. The $GW$-corrected charge transition energies for As$_{Ga}$ are found slightly below midgap for the +2/+1 transition and close to the experimental CBM for the +1/0 transition. In all cases, the magnitude of the relaxation energy $\Delta$ reflects the degree of structural rearrangement between the differing charge states; the $\Delta$ term averages -97 meV for As$_{Ga}$ whereas for Ga$_{As}$ the average is -49 meV.

Comparing the values for $\varepsilon_{q/q'}$ obtained by the DFT+$GW$ approach with those reported in ref. [14] determined from a hybrid DFT approach, reasonable agreement is found for the CTLs occurring near the VBM and the CBM. CTLs occurring deeper in the gap tend to be slightly lower in energy compared to the hybrid DFT results. In general, the bulk defect transition levels calculated by the DFT+$GW$ approximation agree with hybrid DFT results to typically within 100 meV for shallow defect levels ($\varepsilon_{+1/0}$ for V$_{Ga}$, Ga$_{As}$, and As$_{Ga}$) and to within 200 meV for deeper levels ($\varepsilon_{0/-1}$ and $\varepsilon_{+1/0}$ for Ga$_{As}$ and As$_{Ga}$, respectively).

Having focused on the As$_{Ga}$ antisite as a likely candidate for the midgap defects states, previous studies have not reported specific values of CTLs of the Ga$_{As}$ antisite as a function of alloying on the cation sublattice [13, 40]. To examine the influence of local cation disorder, the atoms bonding directly to the Ga$_{As}$ antisite are replaced with either four Ga nearest neighbors or with four In nearest neighbors, recalling that the initial calculation consisted of a configuration with two Ga and two In nearest neighbors denoted as In$_{0.5}$Ga$_{0.5}$-NN. When the defect is surrounded by four nearest neighbor Ga atoms, denoted In$_{0.0}$Ga$_{1.0}$-NN, the CTL $\varepsilon_{0/-1}$(Ga$_{As}$) increases by 60 meV to 0.37 eV. The $\varepsilon_{+1/0}$(Ga$_{As}$) CTL moves to within 20 meV of the VBM for the In$_{0.0}$Ga$_{1.0}$-NN configuration. These changes, relative to the case of In$_{0.5}$Ga$_{0.5}$-NN occur mainly through a decrease in the magnitude of the electron addition energies in the positive charge state $A(q = +1, R_{q=0})$ and $A(q = +1, R_{q=+1})$, on average by 76 meV, while the electron addition energy of the neutral charge state $A(q = 0, R_{q=0})$ decreases by 32 meV. All the relaxation energies ($\Delta$) maintain their values to within 10 meV compared to the values obtained for the In$_{0.5}$Ga$_{0.5}$-NN configuration revealing that the largest change to the electron affinities and ionization potentials are due to the change of the electronegativity of the defect due to local changes in the chemical environment. When the Ga$_{As}$ antisite is bonded to four In atoms In$_{1.0}$Ga$_{0.0}$-NN, the $\varepsilon_{0/-1}$(Ga$_{As}$) CTL increases by approximately 55 meV compared to the In$_{0.5}$Ga$_{0.5}$-NN case resulting in a level at 0.37 eV relative to the VBM. The +1/0 CTL $\varepsilon_{+1/0}$(Ga$_{As}$) = 0.16 eV for In$_{1.0}$Ga$_{0.0}$-NN configuration and is approximately 30 meV lower than for the In$_{0.5}$Ga$_{0.5}$-NN case. While the 0/-1 CTL is approximately equal for the two cases In$_{0.0}$Ga$_{1.0}$ NN and In$_{1.0}$Ga$_{0.0}$ NN, the latter exhibits a slightly increased value of $A(q = 0, R_{q=0})$ of 11 meV compared to In$_{0.5}$Ga$_{0.5}$ NN configuration. The value of $A(q = +1, R_{q=+1})$ decreases by 58 meV relative to the In$_{0.5}$Ga$_{0.5}$-NN bonding configuration and $A(q = +1, R_{q=0})$ increases by 11 meV. Again, all relaxation energies for In$_{1.0}$Ga$_{0.0}$ NN change by less than 10 meV relative to the In$_{0.5}$Ga$_{0.5}$-NN configuration. The largest change in the CTLs for the Ga$_{As}$ antisite as a function of local cation disorder occurs for the +1/0 transition level which lies 0.02 eV above the VBM when bonding to four Ga atoms, compared to 0.19 eV above the VBM when bonding to two In and two Ga atoms. Local cation disorder appears to have a greater effect for donor-like transitions, while acceptor-like transitions change by less than 60 meV.

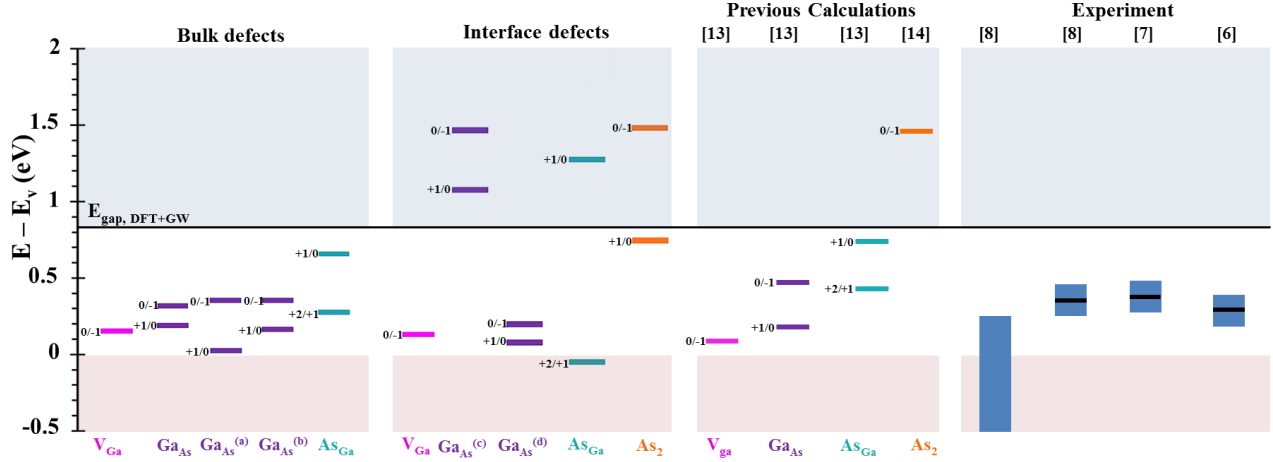

Figure 7 – A comparison of CTLs from this work for bulk (first panel) and interface defects (second panel), previous theoretical work [13,14] (third panel and as also depicted in fig. 1), and experimentally extracted defect states [6,7,8] (fourth panel, levels are shown as dark lines and the broadening/distributions are approximated by the blue rectangles). A band gap of 0.84 eV has been calculated for pristine bulk In$_{0.53}$Ga$_{0.47}$As within the DFT+$GW$ approach using the converged parameters discussed in section II.B. Ga$_{As}$(a) and Ga$_{As}$(b) refer to the Ga antisite bonding to 4 Ga (a), and bonding to 4 In atoms (b). Ga$_{As}$(c) is the Ga antisite calculated in the interface model and bonding directly to the oxide, while Ga$_{As}$(d) is the Ga antisite located 2 monolayers away from the oxide. In the third panel V$_{Ga}$, Ga$_{As}$, As$_{Ga}$ refer to bulk defects, whereas the As$_2$ defect level is for an interface model.

The effect of changes to the cation sublattice alloy configuration has consequences for the donor-like feature of the experimentally observed $D_{it}$ distribution [8]. Variations in the local cation distribution can shift donor-like transition states by up to 200 meV while the CTLs remain within the lower half of the band gap. Therefore, one could conclude that the $D_{it}$ feature occurring below midgap and extending lower to the valence band may have broadened contributions from the variable bonding arrangements due to the random cation alloy; this point will be examined when comparing the experimental broadening of the CTLs.

## B. Interface defects

### 1. Structural properties

The influence of the semiconductor/oxide interface In$_{0.53}$Ga$_{0.47}$As/Al$_2$O$_3$ on the energetics of defects formed near the interface is examined for V$_{Ga}$, Ga$_{As}$ and As$_{Ga}$. The effect on defects directly bonded with atoms forming the oxide layer is also considered. An arsenic-terminated (100) surface bonding to Al$_2$O$_3$ is considered leading to the possibility for the formation of arsenic dimers As$_2$ directly at the interface. The energetics of these interfacial defects are also calculated.

The As$_2$ defect with a configuration as depicted in fig. 8 (a) is constructed following the model of Miceli and Pasquarello [14]. In the model, displacement of the oxide atoms in the (110) direction allow for the formation of two As-As and two Al-Al bonds. An O atom is inserted between each of the two Al-Al bonds and only one of the As-As bonds (see fig. 8 (a)) leaving a single As-As bond at the semiconductor/oxide interface. Relaxing this defect in the neutral charge state, $q = 0$, results in a As-As bond length of 2.56 Å. Charging to $q_{-1}$ increases the As-As bond length to 2.98 Å. This is in agreement with Miceli and Pasquarello who reported an As-As bond length of 2.56±0.01 Å for $q_0$, which increased to 2.95±0.03 Å following energy minimization with respect to the atomic positions in the negative charge state [14]. Charging to $q_{+1}$ results in a small reduction to 2.54 Å in the As-As bond length compared to the $q_0$ case. Thus a larger energy difference is anticipated for the neutral to negative relaxation energy $\Delta(R_{-1}, R_0, q_{-1})$ than for the neutral to positive relaxation energy $\Delta(R_{+1}, R_0, q_{+1})$.

The V$_{Ga}$ model is created by removing a single Ga atom located a monolayer below the oxide, this is the cation layer closest to Al$_2$O$_3$ but not directly bonding to the oxide in this particular model of the interface. All remaining atoms are allowed to relax about the vacancy site. Similar the vacancy model in the bulk, removal of the Ga atom results in an inward movement of the atoms nearest the vacancy by an average of 0.43 Å relative to their positions prior to creation of the vacancy site. Relaxation in the positive charge state results in a slight outward movement of the atoms surrounding the vacancy by 0.01 Å, while relaxation in the negative charge state results in an inward movement of atoms towards the vacancy shown in fig. 8(b) by 0.03 Å. The degrees of atomic displacements are again in proportion to the magnitudes of charge state relaxation.

The $Ga_{As}$ defect in fig. 8(c) bonds directly to the oxide at the interface. For the neutral defect, the defect site is displaced towards the oxide relative to the antisite position prior to performing a geometry relaxation. The defect moves away from Al atoms in the oxide and towards the nearest hydroxyl group -OH for the relaxed configuration; the relaxed $Ga_{As}$-OH and $Ga_{As}$-Al separations are 2.12 Å and 2.59 Å respectively, compared to the corresponding As-OH and As-Al distances prior to creating the antisite defect which average to 3.42 Å and 2.42 Å, respectively. In analogy to the bulk $Ga_{As}$ antisite, the surface $Ga_{As}$ antisite moves towards the nearest Ga neighbor and away from the nearest In neighbor relative to the pristine simulation cell in the absence of the antisite. Relative to the bulk defect, the relaxed surface $Ga_{As}$-Ga bond length is 0.07 Å less than the corresponding surface As-Ga bond length with a 0.08 Å reduction for the bulk $Ga_{As}$ case. The relaxed surface $Ga_{As}$-In distance increases by 0.16 Å relative to the corresponding surface As-In separation compared to only a 0.04 Å increase of the same bond length for the bulk case. Charging to $q_{+1}$ results in a further contraction of the $Ga_{As}$-OH distance to 2.09 Å, while the $Ga_{As}$-Ga and $Ga_{As}$-In separations both increase by an average of 0.02 Å relative to the neutral defect. Adding an electron to the neutral state to bring the charge state to $q_{-1}$ results in a $Ga_{As}$-OH separation of 2.13 Å, a slight increase relative to $q_0$ state. The $Ga_{As}$-Ga and $Ga_{As}$-In separations both decrease by an average of 0.02 Å relative to $q_0$ geometry; *i.e.* the $q_{-1}$ relaxation is almost the same magnitude but opposite in direction compared to $q_{+1}$. These relaxations occur mainly through a movement of the defect center towards a -OH and away from the neighboring cations for $q_{+1}$, and away from -OH and towards the cations for $q_{-1}$.

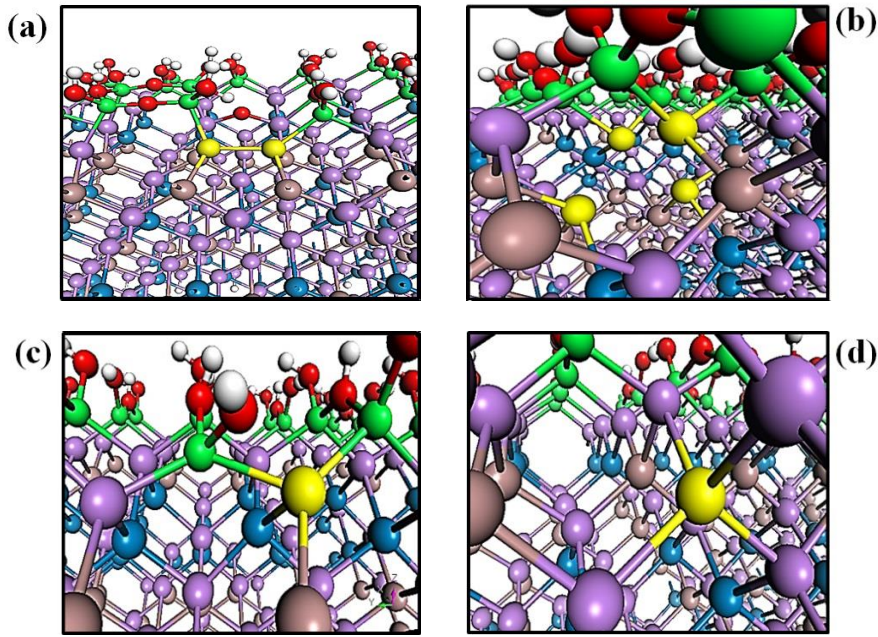

Figure 8 - Relaxed structures of neutral defects ((a) $As_2$, (b) $V_{Ga}$, (c) $Ga_{As}$, (d) $As_{Ga}$) in (100) InGaAs:$Al_2O_3$. Blue, brown and, pink spheres represent In, Ga, and As, resp. Red, white and green spheres represent O, H, and Al, resp. Defect atoms are highlighted in yellow in each image.

The $As_{Ga}$ antisite at the surface is depicted in fig. 8(d) and is not bonding directly to the oxide as the surface of the semiconductor layer is cleaved at a plane of As atoms. Hence as the nearest neighboring bonds are similar to the bulk $As_{Ga}$ defect in terms of structural rearrangement. The $As_{Ga}$-As bonds are on average 0.12 Å greater than the Ga-As bonds in the simulation cell in the absence of the defect. Removing an electron and relaxing in the positive charge state results in 0.02 Å reduction in the $As_{Ga}$-As bond lengths relative to $q_0$, and removing another electron to bring the charge state to $q_{+2}$ yields a further 0.02 Å shortening of $As_{Ga}$-As bond lengths relative to $q_{+1}$.

## 2. CTLs for Interface Defects

For the arsenic dimer surface defect, CTLs calculated using hybrid functional methods as reported in ref. [14] obtain an energy $\varepsilon_{0/-1} = 0.61$ eV above the CBM for the transition level, a result found to be largely independent of the local distribution of indium and gallium atoms on the cation sublattice. Hence the conclusion reached is that the As$_2$ defect is not a likely candidate for creating midgap $D_{it}$ states. However it is noted that the As$_2$ defect is within the GaAs band gap, and is considered a candidate defect for the generally observed inability to accumulate GaAs MOS systems with oxide layers deposited by atomic layer deposition, The $\varepsilon_{0/-1}$ and $\varepsilon_{+1/0}$ transition levels for the arsenic dimer are re-calculated using the DFT+$GW$ approach leading to an electron affinity of $A(q_0, R_0) = 2.16$ eV for the neutral charge state is obtained. The relaxation energy in the negative charge state is calculated in the usual manner

$$\Delta(R_{-1}, R_0, q_{-1}) = E(q_{-1}, R_{-1}) - E(q_{-1}, R_0) \tag{8}$$

leading to $E_{form}(q_{-1}) = A(q_0, R_0) + \Delta(R_{-1}, R_0, q_{-1}) = 1.87$ eV. The neutral charge state is formed by adding an electron to the positive state and subsequently relaxing, yielding $E_{form}(q_0) = 0.38$ eV. For the positive charge state the formation energy is $E_{form}(q_{+1}) = -0.34$ eV. The +1/0 CTL is obtained by taking the difference between $E_{form}(q_0)$ and $E_{form}(q_{+1})$ when the Fermi levels between the two charged states held equal

$$\varepsilon_{+1/0} = \frac{E_{form}(q_0) - E_{form}(q_{+1})}{q_{+1} - q_0}, \tag{9}$$

yielding a positive to neutral CTL of $\varepsilon_{+1/0} = 0.71$ eV relative to the VBM, and similarly

$$\varepsilon_{0/-1} = \frac{E_{form}(q_{-1}) - E_{form}(q_0)}{q_0 - q_{-1}} \tag{10}$$

results in a CTL $\varepsilon_{0/-1} = 0.65$ eV above the theoretical bulk CBM (bulk band gap calculated from DFT+$GW$ = 0.84 eV) in reasonable agreement with the results from the calculations using hybrid functional approximations to DFT reported by Miceli and Pasquarello in ref. [14].

For the Ga$_{As}$ antisite bonding directly to Al$_2$O$_3$, CTLs of $\varepsilon_{0/-1}$= 0.61 eV and $\varepsilon_{+1/0}$= 0.22 eV above the CBM are obtained. Thus, as in the case of the dimer which is likewise occurring directly at the In$_{0.53}$Ga$_{0.47}$As/Al$_2$O$_3$ interface, this antisite defect exhibits CTLs residing well above the conduction band energy of In$_{0.53}$Ga$_{0.47}$As and hence is not expect to play a role in the experimental midgap $D_{it}$. For the defects studied to date, defects bonding directly with the oxide tend to have the CTLs lying much higher in energy due to the distinctly different local chemical bonding.

A V$_{Ga}$ defect created in the Ga layer separated from the oxide layer by the single plane of surface As atoms which terminate the semiconductor slab yields CTLs which are relatively close to the values obtained for the same defect in a bulk environment, but with a slightly lower energy level with $\varepsilon_{0/-1} = 0.14$ eV relative to the VBM, significantly lower than the experimentally reported midgap states but certainly possibly contributing to the high density of defects found near the VBM.

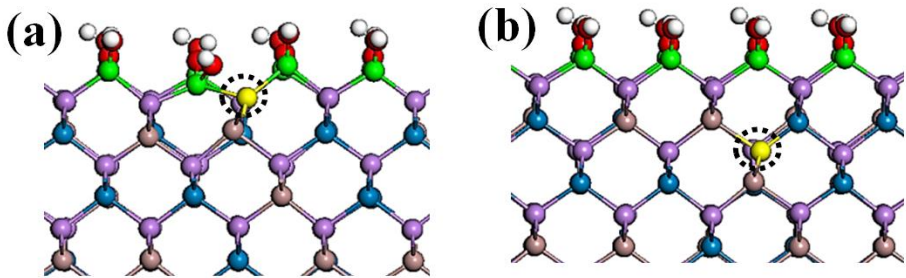

Figure 9 - (a) Ga$_{As}$ anti-site located on the atomic layer which binds to Al$_2$O$_3$. (b) Ga$_{As}$ anti-site located 1 As layer "down" from the oxide. Blue, brown and, pink spheres represent In, Ga, and As, respectively. Red, white and green spheres represent O, H, and Al, respectively. The Ga$_{As}$ anti-site defect is highlighted yellow in both (a) and (b).

Turning now to the As$_{Ga}$ antisite reveals the same qualitative picture as the Ga$_{As}$ antisite and the As$_2$ dimer; defects in close proximity to the oxide, either bonding directly to or up a monolayer away from the oxide, do not exhibit charge transition levels which match the energies corresponding to the distribution of midgap interface states extracted from CV measurements. In the case of the As$_{Ga}$ antisite in close proximity to the oxide, the defect does not bond directly to Al$_2$O$_3$ in the As-terminated (100) In$_{0.53}$Ga$_{0.47}$As interface model as can be seen in fig. 8(d). For this configuration, the CTLs are found to be $\varepsilon_{+2/+1} = -0.06$ eV relative to the VBM and $\varepsilon_{+1/0} = 0.43$ eV relative the CBM, or in other words within the valence and conduction bands respectively and do not give rise to states in the semiconductor band gap. The shift in the CTLs between the bulk- and the As$_{Ga}$ configuration occurring near the interface are found to be significantly smaller than the CTL shifts for the bulk- and surface-model of the Ga$_{As}$ antisite bonding directly to the oxide. However, the latter site has a different local chemistry due its direct bonding to oxide atoms and the larger associated differences in the CTLs relative to bulk defects are not unexpected. If a Ga$_{As}$ anti-site is introduced into the arsenic layer nearest the interface but not bonding to the oxide as can be seen in fig. 9, the CTLs are much nearer to the corresponding values occurring in the bulk with a value of $\varepsilon_{0/-1} = 0.2$ eV relative to the VBM, although still lower in energy by 0.11 eV relative to the same transition in the bulk. A similar shift towards the VBM is also found for the positive to neutral CTL of the Ga$_{As}$ antisite in proximity to the interface with a value of $\varepsilon_{+1/0} = 0.08$ eV for this case. While these values are much closer to the corresponding values in the bulk case, these states are significantly lower in energy than the midgap $D_{it}$ states extracted from measurement [7,8,17,18]. However, clearly these states can be associated with the large defect density seen near the VBM in most experiments. The calculations also indicate that bonding to the oxide plays a significant role in the position of defects states and that it is unlikely, for the defects considered, that bonding directly to oxide have a significant role in introducing defect levels within the band gap of In$_{0.53}$Ga$_{0.47}$As.

*1. Defect formation energies– Bulk versus interface*

A comparison of the formation of interface defects with the bulk counterparts is discussed below. Their formation energies are tabulated in table I.

| Defect | $E_{form}$, bulk | $E_{form}$, interface |
|---|---|---|
| V$_{Ga}$ | 6.94 | 7.09 |
| Ga$_{As}$ | 2.86 | 2.91 |
| Ga$_{As}$ - Al$_2$O$_3$ | N/A | 2.74 |
| As$_{Ga}$ | 1.89 | 1.71 |

Table I - Formation energies of bulk and interface defects calculated using eq. 1. Ga$_{As}$ - Al$_2$O$_3$ refers to the Ga$_{As}$ antisite bonding directly to Al$_2$O$_3$. The formation energy of the Ga$_{As}$ antisite calculated within the surface model but located two monolayers away (see fig. 9) from the oxide is within 50 meV of the bulk counterpart. For V$_{Ga}$ and As$_{Ga}$ calculated in the interface model, these defects bond to the As layer that bonds to Al$_2$O$_3$. All energies in electron volts (eV).

An observation for antisite stability is that bonding directly to the oxide, or bonding to the As layer that bonds to the oxide, results in less energy required to form a given defect than the energy required to form it in the bulk. However, if the antisite is two monolayers or more from the oxide, its formation energy is already within 50 meV of the bulk counterpart. Regarding the vacancy, bulk V$_{Ga}$ defects (relaxed structure shown in fig. 7 (a)) are significantly less stable than the antisites, a trend which has been observed in previous works [13,19,20,41] and continues to hold for defects formed close to the interface. The Ga vacancies near the interface are less stable than their bulk counterparts. The consistent trend of higher formation energies for V$_{Ga}$ compared to other defects implies the reduced concentration of this defect for both bulk and surface variants of this defect center - although this comment is predicated upon an equilibrium argument. For the non-equilibrium conditions that occur during growth, the formation energies are only suggestive of the probability at which various defects can be formed. For the Ga$_{As}$ anti-site bonding directly to the oxide, the formation energy is 0.12 eV lower than the Ga$_{As}$ anti-site in the bulk, and the latter has a formation energy that is 0.97 eV higher compared to the bulk As$_{Ga}$ antisite.

An increase of 0.17 eV in $E_{form}$(Ga$_{As}$) is calculated when this antisite is moved two monolayers away from the semiconductor/oxide interface; refer to fig. 9 for the configuration. This results in an overall decrease in stability of only 50 meV relative to the formation energy in the bulk. Hence even when situated within a few atomic distances from the oxide, the stability of the defect effectively resembles that of the bulk defect. However, again it is noted, the formation energetics do not accurately reflect the energies of the surface during growth and the growth kinetics can alter the picture suggested by the formation energies for defects calculated relative to an ideal interface. Comparing $E_{form}$(As$_{Ga}$) between the bulk and oxide-terminated surface cases, the As$_{Ga}$ antisite exhibits a 0.18 eV increase in the formation energy when the defect is moved from the surface towards the bulk chemical environment.

## IV. Alloy broadening

The shift in CTLs with respect to a large range of defect formation conditions has been considered. These include the cases in which the defect is formed at the interface, near the interface, or in a more bulk like environment. Excluding the large changes in CTLs due a defect directly bonding to the oxide, the shift in energies due to 'proximity' broadening is found to be on the order of 100 meV (see fig. 7 and compare Ga$_{As}$ with Ga$_{As}$$^{(d)}$, the +/0 and 0/- transitions each differ by 110 meV). In addition, for defects formed on the anion sublattice and bonding to the random alloy of In and Ga atoms on the cation sublattice, there is a different local chemical environment based on the specific local distribution of group III atoms. The effect of the random nature of the cation alloy results in an additional broadening for some defects referred to as alloy broadening. To a first approximation, the alloy broadening is ascribed to the nearest neighbor bonds between an anion antisite or vacancy to the cation sublattice. Hence the effect of alloy broadening is anticipated to be small for the As$_{Ga}$ antisite, as there are no nearest neighbor bonds to the group III atoms. Conversely, for an anion situated defect, there is a different local chemical environment due to bonding to the cation sublattice. Here the effect of alloy broadening on the Ga$_{As}$ antisite is considered in detail.

When extracting interface state density profiles from measured CV or conductance-voltage (GV) characteristics, there is an inherent thermal broadening of the actual interface state density distribution due to the occupation of states by the Fermi-Dirac distribution and the method of extraction. This is best illustrated by considering the case of a monoenergetic defect level in an MOS system for a CV measurement and an interface state density profile extraction at room temperature. Due to the finite temperature, a percentage of the interface states are occupied when the Fermi level is below the interface state energy, and as a consequence of the Fermi-Dirac distribution the monoenergetic defect level results in a broad feature on the CV or GV response. When extracting the interface state density profile at each gate voltage (and corresponding surface potential), the difference in the high and low frequency CV characteristics, or the 'stretch out' of the high frequency characteristics, is attributed to an interface state concentration in units of cm$^{-2}$eV$^{-1}$ at that voltage (or surface potential). As a consequence, a monoenergetic level is extracted as a broad feature in the energy gap, and it can easily be shown that the energy distribution of the extracted interface state density profile resulting from a monoenergetic level is precisely the derivate of the Fermi Dirac distribution with respect to energy, at the temperature of the measurement and the interface state density extraction process. From this effect, the minimum thermal broadening for a CTL for measurements made at T = 300 K yields a minimum peak width of 91 meV for full width at half maximum (FWHM). Hence each of the CTLs present in fig. 7 would have a minimum broadening equal to the theoretical lower limit for the thermal broadening of 91 meV. For Ga$_{As}$ the CTLs are re-calculated with either 4 In nearest neighbors, 2 In and 2 Ga nearest neighbors, or 4 Ga nearest neighbors. The energetics for the sites with either 3 In and 1 Ga nearest neighbors or 1 In and 3 Ga nearest neighbors are interpolated. The Ga$_{As}$(+/0) and Ga$_{As}$(0/-) antisite CTLs for each local alloy composition is broadened by the minimum theoretical thermal peak width, and each peak is weighted by the distribution of a random alloy on the cation sublattice. The different CTLs are then summed to give the net effect of the thermal and alloy broadening on the Ga$_{As}$ antisite CTLs. The resulting prediction for the minimum broadening, excluding the proximity broadening, is shown in fig. 10. The net result of accounting for both charge transitions in all 5 local alloy compositions with thermal broadening is shown in fig. 10 (c). It is interesting to note the position of the valley between the two main peaks in fig. 10 (c) (denoted by the dashed arrow). The $D_{it}$ profile reported in ref. [8] exhibits a valley between the midgap $D_{it}$ peak and the broad $D_{it}$ feature extending into the valence band at a similar position in the band gap - approximately 250 meV above the VBM. The similarity of the broadened defect level profile of the Ga$_{As}$ antisite compared to the experimental $D_{it}$ profile suggests that the Ga$_{As}$ should not be ruled out as a midgap $D_{it}$ candidate.

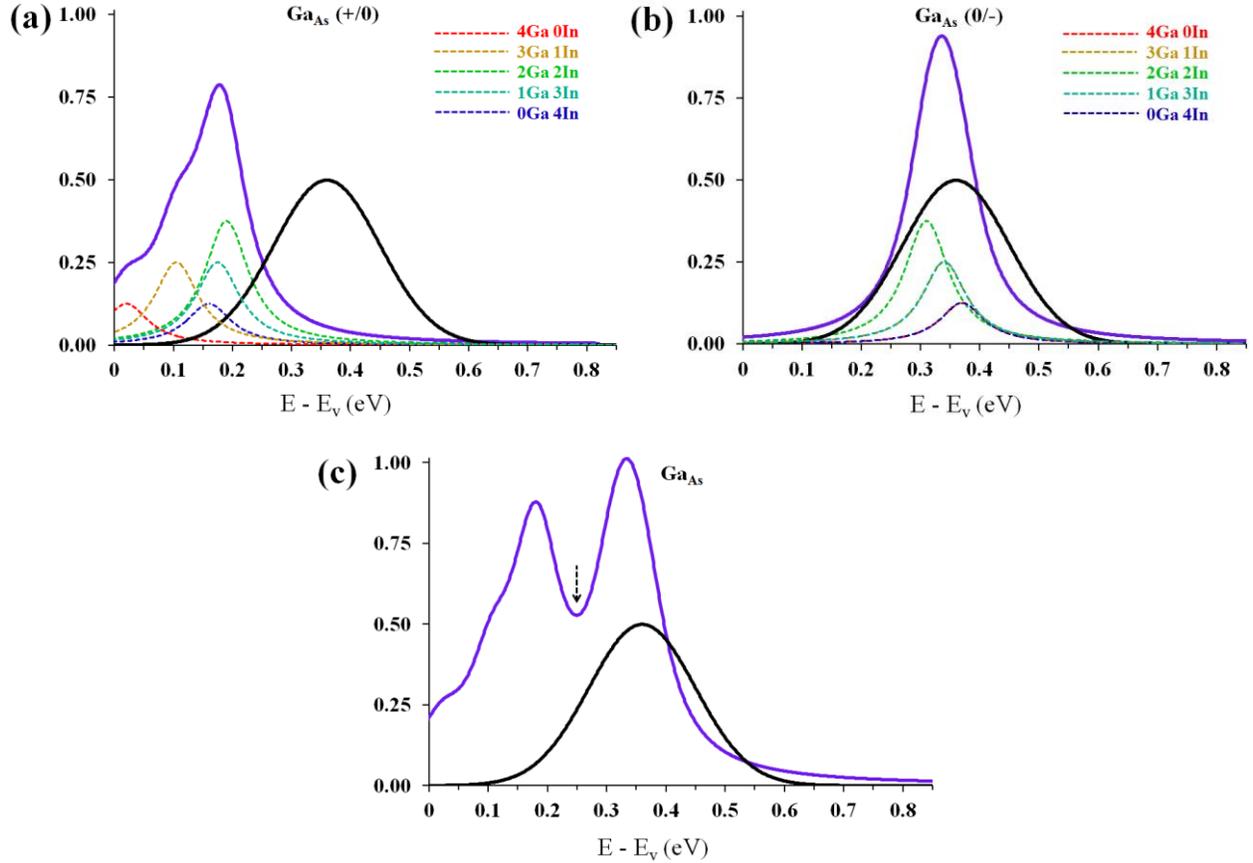

Figure 10 – The effect of thermal and alloy broadening on the (a) $Ga_{As}(+/0)$ and (b) $Ga_{As}(0/-)$ CTLs. The solid black curve in (a), (b), and (c) gives the experimentally extracted midgap defect peak from ref. [8]. The individual dashed peaks are for the CTLs calculated with a different distribution of cation nearest neighbors as identified in the legend within each figure. The solid purple curves in (a) and (b) are the weighted sum for the CTLs due to the alloy composition. The overall effect of this defect is depicted in (c), in which the solid purple curve is the sum of the purple curves from (a) and (b). The valley between the two main peaks is denoted by a dashed arrow, located at approximately 250 meV above the VBM.

Taking into consideration the results of this analysis, the additional energy shifts to the CTLs due to defects being formed near the semiconductor/oxide interface, and the error inherent in the calculations, we infer the following. Although the $As_{Ga}(+2/+1)$ CTL has been identified as a candidate for the midgap states, the $Ga_{As}$ antisite should not be ruled out as an additional strong candidate for generating the defect states at midgap in CV measurements. There have been two primary considerations that have previously focused attention on the $As_{Ga}$ antisite. First is the experimental finding that the integrated charge across the energy gap is positive [8], and the second is that the calculated formation energy for the $Ga_{As}$ antisite is on the order of 1 eV higher than the formation energy of the $As_{Ga}$ antisite.

The first point does not rule out the $Ga_{As}$ antisite as giving rise to the midgap states based on the experimental data. CV measurements indicate that the net charge due to the defect states integrated over the energy gap is positive. However, the midgap states form a relatively small peak and there is a much higher defect density in the gap but nearer to the valence band edge. The overall effect of a high density of positively charged defects giving rise to defect levels near the band edge maximum can compensate for the effect of a smaller population of negatively charged defects yielding midgap states. Hence the overall charge contribution due to all defects does not *a priori* eliminate the possibility that the $Ga_{As}$ antisite generates midgap levels.

The second point is based on the formation energies of the antisite defects. An analysis based upon formation energies relies on an equilibrium process for the formation of defects. The non-equilibrium processes occurring during the growth of a material such as $In_{0.53}Ga_{0.47}As$ cannot be reduced to a set of processes described by equilibrium thermodynamics. For example, to provide reasonable growth conditions an As-rich supply of carrier gases must be provided to the growth chamber - the ratio of the group V to group III precursors can exceed a factor of a hundred. However, this does not reflect the ratio of group III and V atoms available for growth at the surface of a substrate in the chamber. Hence although the growth conditions are described as 'As-rich', small variations in the growth conditions can lead to either 'As-rich' or 'In/Ga-rich' conditions at the growth front. Hence the possibility of forming either or both $As_{Ga}$ and $Ga_{As}$ antisites during the growth of $In_{0.57}Ga_{0.43}As$.

Hence our analysis leads us to the following conjecture. Some electrical characterizations see a peaked feature in the midgap $D_{it}$ [8,17,18], whereas others do not [42]. Experiments suggest that the midgap states are largely independent of the oxide layer deposited on the $In_{0.53}Ga_{0.47}As$ surface [6-9] and thus the defects are attributed to the semiconductor layer. Hence our analysis suggests that a possible explanation is that there are many defect states providing positive charge with CTLs near the VBM maximum, and these states are observed in most CV measurements. However, if growth conditions are favorable to the formation of antisites, additional features at midgap are observed. If the charge associated with the midgap states increases the overall positive charge contribution of all states in the band gap, then the defects should be identified with $As_{Ga}$ antisites. If the net effect of the midgap states is to reduce the overall positive charge contribution of the states in the band gap (but the integrated charge remains positive), then the midgap states can be associated with $Ga_{As}$ or related defects and the broad distribution of these midgap states can be associated with thermal, alloying and proximity broadening.

## V. Conclusions

Qualitative agreement for energies of CTLs for a range of defects native to $In_{0.53}Ga_{0.47}As$ is found compared to previous computational studies based on hybrid DFT methods. Shallow defect levels agree well between the DFT+$GW$ approach and hybrid DFT. Deeper levels calculated by the DFT+$GW$ approach are predicted to be 100 to 200 meV lower in energy compared to hybrid DFT results for defects occurring near and below mid gap [13]. Good agreement between the methods is also found for the $As_2$ interface defect with both approaches predicting a transition level resonant with the conduction band [14]. Our findings are consistent, to within the accepted theoretical and experimental accuracy, with the conclusions of Komsa and Pasquarello [13] that the $As_{Ga}$ antisite in a bulk-like local chemical environment is a candidate for giving rise to the electrically states below midgap as observed at $In_{0.53}Ga_{0.47}As$/high-$k$ oxide interfaces. However, a detailed investigation of the $Ga_{As}$ antisite reveals that if this defect site is incorporated during kinetic growth conditions, then the CTLs associated with this defect and the effects of alloy broadening and proximity broadening all suggest that this defect level is also compatible with the observation of midgap defect states, and consistent with the broadening associated with this feature. The fact that the midgap states are only observed in some experimental CV measurements may be an indication that these defects are only present in significant concentration under some growth conditions.

The stability of the neutral defect centers in various chemical environments is also evaluated. As is expected due to the loss of bonding energy upon formation of a vacancy, a higher formation energy for the $V_{Ga}$ compared to other defects in the bulk and oxide-terminated surface models is found, suggesting a lower concentration for this defect and hence a relatively small contribution to the $D_{it}$ observed in III-V/high-$k$ oxide MOS devices. The lower formation of the $As_{Ga}$ antisite compared to other defects is observed; the $Ga_{As}$ formation energy is approximately one electron volt higher relative to the $As_{Ga}$ antisite. During As-rich conditions or In/Ga-rich conditions at the surface during growth, it is not clear how to relate equilibrium formation energies to the population of the antisite densities. Hence, although the formation energies give some indication of relative stabilities, the actual relative population of the different defects will depend heavily upon growth conditions.

Comparing the CTLs for bulk-like defects and for defects formed near the semiconductor/oxide interface, summarized in fig. 7, reveals a large range of energies for the various defect levels. The defect levels have been determined for bulk-like environments and these studies have been complemented by a quantitative analysis of the shift in the CTL energies as a function of their distance from the oxide. The calculations show that CTLs return to approximately 100 meV of the corresponding bulk values when situated even a few monolayers from the semiconductor/oxide interface. This is consistent with previous studies in which the defect levels of antisites located at various distances from a GaAs surface were found to return to bulk-like values within a few monolayers from the surface [43]. In our simulations, the $Ga_{As}$ antisite exhibits CTLs which resonate with the $In_{0.53}Ga_{0.47}As$ conduction band when directly bonded to the oxide at the semiconductor surface, however the defect energetics become significantly more like their bulk counterparts even if situated two monolayers away from the semiconductor/oxide interface, being on the order of 0.1 eV of the corresponding bulk values. In addition, variable bonding arrangements due to local alloy disorder may account for the $D_{it}$ feature extending into the valence band [8,44], while defects bonding directly to the oxide have transition levels deep within the conduction band, potentially explaining the increase in $D_{it}$ near and above the $In_{0.53}Ga_{0.47}As$ conduction band edge observed in a number of experimental works [8,18,44]. This is also consistent with recent theoretical results [14,46,47].

III-V oxide interfaces have not received the amount of attention, particularly at the atomistic level, of the silicon-silicon oxide interfaces [48] or silicon-high k interfaces [30], however their increasing technological relevance is generating increased interest in these complex systems. Although there remain clear questions as to the specific nature of the atomic structures giving rise to the interface density of states with the band gap of $In_{0.53}Ga_{0.47}As$/oxide interfaces, it appears that there are a variety of contributions and these contributions can potentially vary due to growth conditions. However, antisites seem to be the most likely candidates for generating midgap states *and* for generating a strong contribution to the high density of defect levels seen at the valence band edge.

## ACKNOWLEDGMENTS

This work was performed in cooperation with the Intel Corporation under the Tyndall-Intel Research Collaboration 2013-2015 and additionally supported by the European Union project DEEPEN funded under NMR-2013-1.4-1 grant agreement number 604416. Additional support was provided through Science Foundation Ireland through the Principal Investigator grant 13/IA/1956. We are grateful to Dr. Rafael Rios for valuable discussions on the implications of this study for transistor engineering and Professor Emanuele Pelucchi for helpful discussions on kinetic growth conditions for compound semiconductors.

## REFERENCES

[1] Theory *Of Defects In Semiconductors*, edited by D. A. Drabold and S. K. Estreicher, (Springer, 2007).
[2] P. Ebert, Curr. Opin. Solid. St. M. **5**, 211 (2001).
[3] K. Cherkaoui, V. Djara, E. O'Connor, J. Lin, M. A. Negara, I. M. Povey, S. Monaghan, and P. K. Hurley, ECS Transactions **45**, 79 (2012).
[4] D. J. Frank, IBM J. Res. Dev. **46**, 235 (2002).
[5] J. A. del Alamo, Nature **479**, 317 (2011).
[6] P. K. Hurley, E. O'Connor, S. Monoghan, R. D. Long, A. O'Mahony, I. M Povey, K. Cherkaoui, J. Machale, A. J. Quinn, G. Brammertz, M. Heyns, S. B Newcomb, V. V. Afanas'ev, A. M Sonnet, R. V. Galatage, M. N. Jivani, E. M. Vogel, R. M. Wallace, M. E. Pemble, ECS Transactions **25**, 113 (2009).
[7] E. O'Connor, B. Brennan, V. Djara, K. Cherkaoui, S. Monoghan, S. B. Newcomb, R. Contreras, M. Milojevic, G. Hughes, M. E. Pemble, R. M. Wallace and P. K. Hurley, J. Appl. Phys. **109**, 024101 (10) (2011).
[8] V. Djara, T. P. O'Regan, K. Cherkaoui, M. Schmidt, S. Monoghan, I.M. Povey, D. O'Connell, M.E. Pemble, P.K. Hurley, Microelectron. Eng. **109**, 182 (2013).
[9] K. Cherkaoui, E. O'Connor, S. Monaghan, R. D. Long, V. Djara, A. O'Mahony, R. Nagle, M. E. Pemble, and P. K. Hurley, Dielectrics for Nanosystems 4: Materials Science, Processing, Reliability, and Manufacturing **28**, 181 (2010).
[10] H. P. Komsa and A. Pasquarello, Appl. Phys. Lett. **97**, 191901 (3) (2010).
[11] H. P. Komsa and A. Pasquarello, Microelectron. Eng. **88**, 1436 (2011).
[12] P. Komsa and A. Pasquarello, Physica B **407**, 2833 (2012).
[13] P. Komsa and A. Pasquarello, J. Phys.: Condens. Mat. **24** (2012).
[14] G. Miceli and A. Pasquarello, Microelectron. Eng. **109**, 60 (2013).
[15] J. Robertson and L. Lin, Appl. Phys. Lett. **99** (2011).
[16] M. Hedström, A. Schindlmayr, G. Schwarz, and M. Scheffler, Phys Rev Lett **97**, 266401 (4) (2006).
[17] P. F. Zhang, R. E. Nagle, N. Deepak, I. M. Povey, Y. Y. Gomeniuk, E. O'Connor, N. Petkov, M. Schmidt, T. P. O'Regan, K. Cherkaoui, M. E. Pemble, P. K. Hurley, R. W. Whatmore, Microelectron. Eng. **88**, 1054 (2011).
[18] A. Ali, H. Madan, S. Koveshnikov, S. Oktyabrsky, R. Kambhampati, T. Heeg, D. Schlom, and S. Datta, IEEE T. Electron Dev. **57**, 742 (2010).
[19] C. G. Van de Walle and J. Neugebauer, J. Appl. Phys. **95**, 3851 (2004).
[20] A. Chroneos, H. A. Tahini, U. Schwingenschlögl, and R. W. Grimes, J. Appl. Phys. **116** (2014).
[21] R. W. Godby, M. Schluter, and L. J. Sham, Phys. Rev. Lett. **56**, 2415 (1986).
[22] J. P. Perdew and M. Levy, Phys. Rev. Lett. **51**, 1884 (1983).
[23] J. P. Perdew and A. Zunger, Phys. Rev. B. **23**, 5048 (1981).
[24] P. Giannozzi, et al., J. Phys.: Condens. Mat **21** (2009).
[25] I. Vurgaftman, J. R. Meyer, and L. R. Ram-Mohan, J. Appl. Phys. **89**, 5815 (2001).
[26] A. Korkin, J. C. Greer, G. Bersuker, V. Karasiev, and R. J. Bartlett, Physical Review B **73**, 165312 (2006).

[27] S. H. Wei, L. G. Ferreira, J. E. Bernard, and A. Zunger, Phys. Rev. B. **42**, 9622 (1990).
[28] A. Zunger, S. H. Wei, L. G. Ferreira, and J. E. Bernard, Phys. Rev. Lett. **65**, 353 (1990).
[29] S. D. Elliott and J. C. Greer, Appl. Phys. Lett. **98** (2011).
[30] S. Monaghan, J. C. Greer and S. D. Elliott, Physical Review B **75**, 245304 (2007)
[31] A. Zunger, Appl. Phys. Lett. **83**, 57 (2003).
[32] C. Freysoldt, B. Grabowski, T. Hickel, J. Neugebauer, G. Kresse, A. Janotti, and C. G. Van de Walle, Rev. Mod. Phys. **86**, 253 (2014).
[33] P. Rinke, A. Janotti, M. Scheffler, and C. G. Van de Walle, Phys. Rev. Lett. **102**, 026402 (2009).
[34] A. Marini, C. Hogan, M. Grüning, and D. Varsano, Comput Phys Commun **180**, 1392 (2009).
[35] W. G. Aulbur, L. Jonsson, and J. W. Wilkins, Solid State Phys. **54**, 1 (2000).
[36] M. S. Hybertsen and S. G. Louie, Phys. Rev. B. **34**, 5390 (1986).
[37] A. Marini, G. Onida, and R. Del Sole, Phys. Rev. Lett. **88**, 016403 (4) (2002).
[38] H. C. Lin, W. E. Wang, G. Brammertz, M. Meuris, and M. Heyns, Microelectron. Eng. **86**, 1554 (2009).
[39] G. Schwarz, A. Kley, J. Neugebauer, and M. Scheffler, Phys. Rev. B. **58**, 1392 (1998).
[40] S. T. Murphy, A. Chroneos, R. W. Grimes, C. Jiang, and U. Schwingenschlögl, Phys. Rev. B. **84**, 184108 (2011).
[41] H. A. Tahini, A. Chroneos, S. T. Murphy, U. Schwingenschlögl, and R. W. Grimes, J. Appl. Phys. **114** (2013).
[42] G. Brammertz, H. C. Lin, M. Caymax, M. Meuris, M. Heyns, and M. Passlack, Appl. Phys. Lett. **95** (2009).
[43] R. B. Capaz, K. Cho, and J. D. Joannopoulos, Phys. Rev. Lett. **75**, 1811 (1995).
[44] G. Brammertz, H. C. Lin, K. Martens, A. Alian, C. Merckling, J. Penaud, D. Kohen, W.-E Wang, S. Sioncke, A. Delabie, M. Meuris, M. Caymax, M. Heyns, ECS Transactions **19**, 375 (2009).
[46] L. Lin and J. Robertson, J. Vac. Sci. Technol. B. **30** (2012).
[47] Y. Z. Guo and J. Robertson, Microelectron. Eng. **109**, 274 (2013).
[48] A. Korkin, J. C. Greer, G. Bersuker, V. Karasiev, and R. J. Bartlett, Physical Review B **73**, 165312 (2006)